\documentclass[reprint,amssymb,fleqn, amsmath, aps, superscriptaddress, showpacs, nobibnotes, nofootinbib, pra]{revtex4-1}
\usepackage[T1]{fontenc}
\usepackage[utf8]{inputenc}
\usepackage{amssymb}
\usepackage{amsmath}
\usepackage{color}
\usepackage{hyperref}
\usepackage{verbatim}
\usepackage{graphicx}
\usepackage{gensymb}
\begin{document}
\title{Charge transfer energy in iridates: a hard x-ray photoelectron spectroscopy study}
\author{D. Takegami}
\affiliation{Max Planck Institute for Chemical Physics of Solids, N{\"o}thnitzer Stra{\ss}e 40, 01187 Dresden, Germany}
\author{D. Kasinathan}
\affiliation{Max Planck Institute for Chemical Physics of Solids, N{\"o}thnitzer Stra{\ss}e 40, 01187 Dresden, Germany}
\author{K. K. Wolff}
\affiliation{Max Planck Institute for Chemical Physics of Solids, N{\"o}thnitzer Stra{\ss}e 40, 01187 Dresden, Germany}
\author{S. G. Altendorf}
\affiliation{Max Planck Institute for Chemical Physics of Solids, N{\"o}thnitzer Stra{\ss}e 40, 01187 Dresden, Germany}
\author{C. F. Chang}
\affiliation{Max Planck Institute for Chemical Physics of Solids, N{\"o}thnitzer Stra{\ss}e 40, 01187 Dresden, Germany}
\author{K. Hoefer}
\affiliation{Max Planck Institute for Chemical Physics of Solids, N{\"o}thnitzer Stra{\ss}e 40, 01187 Dresden, Germany}
\author{A. Melendez-Sans}
\affiliation{Max Planck Institute for Chemical Physics of Solids, N{\"o}thnitzer Stra{\ss}e 40, 01187 Dresden, Germany}
\author{Y. Utsumi}
\altaffiliation[Current address: ]{Institute of Physics, Bijeni\v{c}ka 46, 10000 Zagreb, Croatia}
\affiliation{Max Planck Institute for Chemical Physics of Solids, N{\"o}thnitzer Stra{\ss}e 40, 01187 Dresden, Germany}
\author{F. Meneghin}
\affiliation{Max Planck Institute for Chemical Physics of Solids, N{\"o}thnitzer Stra{\ss}e 40, 01187 Dresden, Germany}
\affiliation{Politecnico di Milano, Piazza Leonardo da Vinci, 32, 20133 Milano, Italy}
\author{T. D. Ha}
\affiliation{Max Planck Institute for Chemical Physics of Solids, N{\"o}thnitzer Stra{\ss}e 40, 01187 Dresden, Germany}
\affiliation{Department of Electrophyiscs, National Chiao Tung University, 1001 Ta Hsueh Road, 30010 Hsinchu, Taiwan.}
\author{C. H. Yen}
\affiliation{Max Planck Institute for Chemical Physics of Solids, N{\"o}thnitzer Stra{\ss}e 40, 01187 Dresden, Germany}
\affiliation{Department of Physics, National Tsing Hua University, 101 Kuang Fu Road, 30013 Hsinchu, Taiwan}
\author{K. Chen}
\affiliation{Institute of Physics II, University of Cologne, Z{\"u}lpicher Stra{\ss}e 77, 50937 Cologne, Germany}
\author{C. Y. Kuo}
\affiliation{Max Planck Institute for Chemical Physics of Solids, N{\"o}thnitzer Stra{\ss}e 40, 01187 Dresden, Germany}
\affiliation{National Synchrotron Radiation Research Center (NSRRC), 101 Hsin-Ann Road, 30076 Hsinchu, Taiwan}
\author{Y. F. Liao}
\affiliation{National Synchrotron Radiation Research Center (NSRRC), 101 Hsin-Ann Road, 30076 Hsinchu, Taiwan}
\author{K. D. Tsuei}
\affiliation{National Synchrotron Radiation Research Center (NSRRC), 101 Hsin-Ann Road, 30076 Hsinchu, Taiwan}
\author{R. Morrow}
\affiliation{Leibniz Institute for Solid State and Materials Research IFW Dresden, Helmholtzstra{\ss}e 20, 01069 Dresden, Germany}
\author{S. Wurmehl}
\affiliation{Leibniz Institute for Solid State and Materials Research IFW Dresden, Helmholtzstra{\ss}e 20, 01069 Dresden, Germany}
\author{B. B{\"u}chner}
\affiliation{Leibniz Institute for Solid State and Materials Research IFW Dresden, Helmholtzstra{\ss}e 20, 01069 Dresden, Germany}
\affiliation{Institut f{\"u}r Festk{\"o}rperphysik , Technische Universit{\"a}t Dresden, 01062 Dresden, Germany}
\author{B. E. Prasad}
\affiliation{Max Planck Institute for Chemical Physics of Solids, N{\"o}thnitzer Stra{\ss}e 40, 01187 Dresden, Germany}
\author{M. Jansen}
\affiliation{Max Planck Institute for Solid State Research, Heisenbergstra{\ss}e 1, 70569 Stuttgart,Germany}
\author{A. C. Komarek}
\affiliation{Max Planck Institute for Chemical Physics of Solids, N{\"o}thnitzer Stra{\ss}e 40, 01187 Dresden, Germany}
\author{P. Hansmann}
\affiliation{Max Planck Institute for Chemical Physics of Solids, N{\"o}thnitzer Stra{\ss}e 40, 01187 Dresden, Germany}
\affiliation{Department of Physics, University of Erlangen - Nuremberg, 91058 Erlangen, Germany}
\author{L. H. Tjeng}
\affiliation{Max Planck Institute for Chemical Physics of Solids, N{\"o}thnitzer Stra{\ss}e 40, 01187 Dresden, Germany}

\date{\today}

\begin{abstract}
We have investigated the electronic structure of iridates in the double perovskite crystal structure containing either 
Ir$^{4+}$ or Ir$^{5+}$ using hard x-ray photoelectron spectroscopy. The experimental valence band 
spectra can be well reproduced using tight binding calculations including only the Ir $5d$, O $2p$ and O $2s$ 
orbitals with parameters based on the downfolding of the density-functional band structure results. We found 
that regardless of the A and B cations, the A$_2$BIrO$_6$ iridates have essentially zero O $2p$ to Ir $5d$ charge 
transfer energies. Hence, double perovskite iridates turn out to be extremely covalent systems with the consequence being that the magnetic 
exchange interactions become very long-ranged, thereby hampering the materialization of the long-sought Kitaev 
physics. Nevertheless, it still would be possible to realize a spin-liquid system using the iridates with a proper 
tuning of the various competing exchange interactions.
\end{abstract}


\maketitle

\section{Introduction}
Recently, the class of iridium oxide materials has attracted tremendous interest due to the expectation for exotic magnetic states which could arise as a consequence of the interplay between the strong spin-orbit coupling, crystal field, and Coulomb interactions. 
For iridates with the formal Ir$^{4+}$ valency and a locally cubic coordination, it has been proposed that 
the $t_{2g}$ states split into a fully filled $j_{eff}=3/2$ band and a half filled $j_{eff}=1/2$ band, leading 
to a pseudospin $J_{eff}=1/2$ Mott insulating state as a ground state in the Ir$^{4+}$ ions \cite{Kim2008}. 
Such a $J_{eff}=1/2$ ground state would potentially realize the paradigmatic Kitaev model 
\cite{Kitaev2006,Jackeli2009,Chaloupka2010} which led to an extensive work on various candidate
materials \cite{Winter2017,Takagi2019}. However, it turned out that perturbations from the ideal cubic coordination 
or longer range hopping can introduce other types of intersite exchange interactions 
masking the long-sought pure Kitaev phenomenon \cite{Winter2016}.

For compounds with the formal Ir$^{5+}$ valency, the analogous picture leaves the $j_{eff}=3/2$ bands fully 
filled while the $j_{eff}=1/2$ doublet remains empty, producing a Van Vleck singlet ground state with $J_{eff}=0$ \cite{Wolff2017,Prasad2018,Wolff2019}. Interestingly, a theoretical study presented yet the possibility of the presence 
of excitonic magnetism in such systems \cite{Khaliullin2013}, and some experimental studies reported also unusual 
magnetic behavior in Ir$^{5+}$ compounds such as Sr$_2$YIrO$_6$ and Ba$_2$YIrO$_6$ \cite{Cao2014, Terzic2017}. 
The origin of such behaviour and the possibility for the materialization of such exitonic magnetism are still subject 
to debate \cite{Bhowal2015,Pajskr2016,Dey2016,Corredor2017,Kusch2018}.

\begin{figure}
\centering
    \includegraphics[width=0.98\columnwidth]{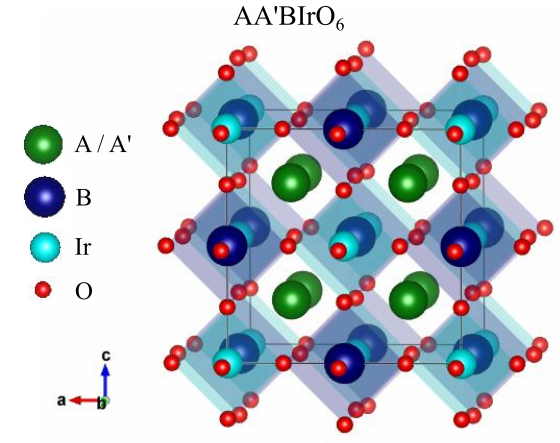}
    \caption{Crystal structure of the double perovskite AA'BIrO$_6$. Red spheres correspond to the position of oxygen atoms, the dark green spheres to the A and A' sites. Cyan and dark blue octahedra correspond to the IrO$_6$ and BO$_6$ octahedra, respectively. }
    \label{fig_0}
\end{figure}

In this context, double perovskite iridates of AA'BIrO$_6$ form{ , the crystal structure of which is displayed in Fig.~\ref{fig_0},} provide a very interesting playground \cite{Vasala2015}
for the search and development of materials with novel magnetic and electric properties. The substitution 
of the A, A' and B sites allows for a tuning of the different competing interactions, local symmetries and ligand environment 
as well as the oxidation state of the iridium. A significant amount of research has been reported during the last years 
\cite{ Wolff2017, Prasad2018, Wolff2019, Cao2014, Terzic2017, Bhowal2015, Pajskr2016, Dey2016, Corredor2017, 
Kusch2018, Kolchinskaya2012, Cao2013, Laguna-Marco2015, Cook2015, Aczel2016, Fuchs2018, Aczel2019, Agrestini2019}. 
Also, in the double perovskites the Ir-Ir distance is much larger compared to the honeycomb systems which reduces delocalization
(i.e.\ band formation) and leads to better defined local $J_{eff}=1/2$ entities. A new class of face-centered-cubic materials
for Kitaev physics has been suggested based on the double perovskite iridates \cite{Cao2013,Cook2015,Aczel2016,Aczel2019}.

While the larger Ir-Ir distances and the chemical and concomitant structural tunability offered by the double perovskite structure are indeed a step
forward towards the necessary conditions for the realization of the Kitaev model, another issue remains  to be addressed. As it has been argued in the case of Sr$_2$IrO$_4$ \cite{Agrestini2017}, covalency may be expected to be large in iridates generally. The assumption of an effective $J_{eff}=1/2$ state, therefore, may become increasingly questionable for larger hybridization strengths since it is based on an ionic Ir $5d$ $t_{2g}^5$ configuration.

In this article, we present a systematic hard x-ray photoelectron spectroscopy (HAXPES) study of double perovskite 
iridates with Ir$^{4+}$ (La$_2$BIrO$_6$, with B = Mg, Co, Ni, Zi) and Ir$^{5+}$ (Ba$_2$YIrO$_6$, Sr$_2$YIrO$_6$, 
Sr$_2$FeIrO$_6$, Bi$_2$NaIrO$_6$ and SrLaBIrO$_6$, with B= Ni, Zn). With this wide range of compounds 
we study the general features of the electronic structure of the double perovskite iridates. Our main focus is 
on the issue of covalency and less on the aspects that are the result of the small structural variations or the 
magnetic properties of the B site cations. One advantage of using photoelectron spectroscopy %for this study 
in comparison to absorption based spectroscopies is that photoemission has a much higher sensitivity to 
covalency \cite{Degroot1994}. Furthermore, we choose the HAXPES variant of all photoelectron spectroscopic
techniques to exploit its large probing depth and thus to obtain spectra that are representative
of the bulk material. In addition, at high photon energies the photoionization cross-sections of the Ir $5d$ are much larger than those of the other orbitals from lighter elements contributing to the valence band \cite{TRZHASKOVSKAYA200197,TRZHASKOVSKAYA2002257,TRZHASKOVSKAYA2006245}. We thus can expect 
that our spectra directly unveil the Ir 5$d$ contributions \cite{Kahk2014,Yamasaki2014}. The experimental 
data is complemented with \textit{ab-initio} density functional calculations and subsequent downfolding to effective localized Wannier bases, in order to get a quantitative understanding of the local electronic structure of the iridium in the double perovskite iridates. 

\section{Methods}

Single crystals of the corresponding iridium double perovskites La$_2$NiIrO$_6$, La$_2$ZnIrO$_6$, La$_2$MgIrO$_6$, La$_2$CoIrO$_6$ were grown using pre-reacted powders of the targeted double-perovskite composition (see, e.g., Vogl \textit{et al.} \cite{Vogl2018}). About 5~g powders of each precursor were ground and mixed with PbO:PbF$_2$ in a 1:1 mass ratio of flux components. Precursor and flux mixture was put in a Pt crucible, tightly closed with a Pt lid, and heated to 1200$^\circ$C with a dwell time of about 24~h followed by slow cooling with 1.7$^\circ$C/h. After growth, the crystals were mechanically separated from the solidified flux and residual flux was washed off with dilute nitric acid.

All crystals were carefully characterized regarding their structure (by powder x-ray diffraction with STOE STADI laboratory diffractometer (transmission geometry with Mo K$_{\alpha1}$ radiation from a germanium monochromator and a DECTRIS MYTHEN 1K detector) and by single crystal diffractometry), homogeneity and composition (EVO MA 10 (ZEISS) scanning electron microscope with an energy-dispersive X-ray analyzer (OXFORD instruments) and magnetic properties (magnetometry using a Quantum Design MPMS-XL SQUID magnetometer).

Single crystals of Ba$_2$YIrO$_6$, Sr$_2$YIrO$_6$, and Bi$_2$NaIrO$_6$ and polycristalline samples of Sr$_2$FeIrO$_6$, and SrLaBIrO$_6$ with B=(Ni, Zn) were grown following the procedures as described in the literature \cite{Dey2016,Corredor2017,Prasad2018,Page2018,Wolff2017}.

The experiments have been carried out at the Max-Planck-NSRRC HAXPES end-station at the Taiwan undulator beamline BL12XU at SPring-8, Japan. The photon beam was linearly polarized 
with the electrical field vector in the plane of the storage ring (i.e.\ horizontal) and the photon energy was set at about
6.5 keV. An MB Scientific A-1 HE analyzer, mounted horizontally, was used \cite{Weinen2015}. The photoelectrons were collected in the 
direction parallel to the electrical field vector of the photon beam.
{ Measurements with the photoelectrons collected in the perpendicular direction were also performed for La$_2$MgIrO$_6$, Bi$_2$NaIrO$_6$, and Sr$_2$YIrO$_6$ (see Appendix).}
The overall energy resolution was set at around 0.3 eV. 
Clean sample surfaces were obtained by cleaving the samples \textit{in situ} in an ultra-high vacuum preparation 
chamber with a pressure in the $10^{-10}$~mbar range. All measurements were performed at 80 K except for the 
SrLaNiIrO$_6$ and Sr$_2$YIrO$_6$, which were performed at 300 K. 

To compute total and partial (i.e.\ orbitally resolved) single particle density of states (DOS, PDOS), we performed non-spin polarized (scalar relativistic) density functional theory (DFT) calculations within
the local density approximation (LDA) using the full-potential local-orbital (FPLO) code \cite{Koepernik99}.
For the Brillouin zone (BZ) integration we used the tetrahedron method with a  $12 \times 12 \times 12$ $\mathbf{k}$-mesh. { The crystal structures used for the calculations correspond to the experimental room temperature crystal structures reported in the literature \cite{Currie95,Dey2016,Corredor2017,Page2018,Prasad2018,Wolff2017}. In the literature, it is also reported that no significant crystal structure changes occur at low temperature.}

\begin{figure*}
\centering
    \includegraphics[width=1.98\columnwidth]{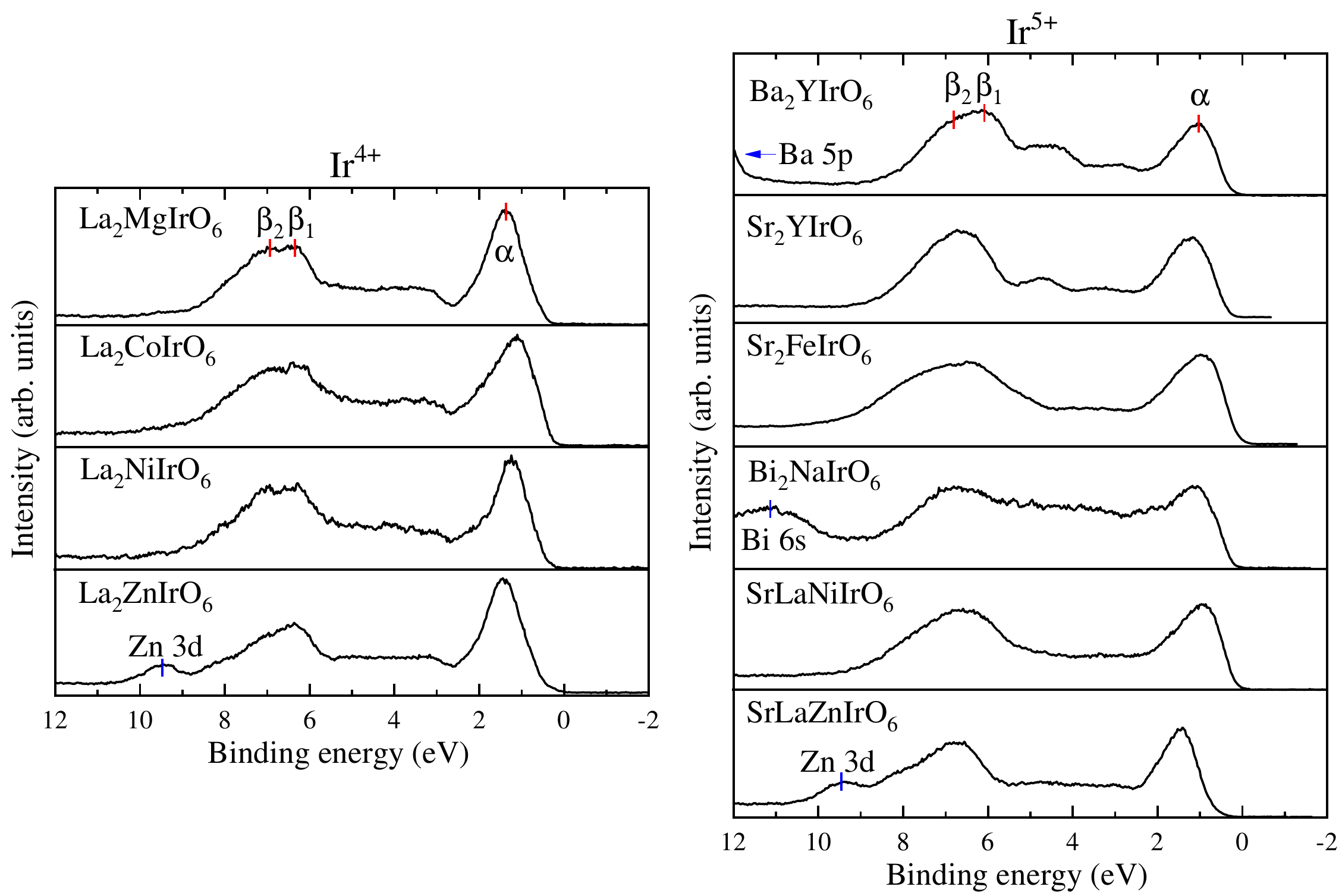}
    \caption{Valence band HAXPES spectra ($h\nu$ = 6.5 keV) of the double perovskite 
     compounds containing formally Ir$^{4+}$ (left panel) and Ir$^{5+}$ (right panel) ions. }
    \label{fig_1}
\end{figure*}

For the derivation of the effective hopping integrals we downfolded to a lattice basis of Wannier functions including O $2p$, $2s$ and  Ir $5d$ orbitals. In order to translate the numerical downfolding results into analytical hopping terms, we mapped the model to a  Slater-Koster tight-binding model on a linear combination of atomic orbitals (LCAO) basis \cite{Slater54}
which consists of O $2p$, $2s$ and  Ir $5d$ orbitals and optimized to match the experimental spectra.

\section{Results}

Figure \ref{fig_1} shows the HAXPES valence band spectra of the studied double perovskite iridates. We can 
observe that all measured samples have no or negligible spectral weight at the Fermi level (zero binding energy),
consistent with their insulating behavior. Remarkably, all spectra look rather similar: there are two main 
features, namely a narrower peak at around 1 eV binding energy (labelled $\alpha$) and a broader structure between 
6 and 8 eV (labelled $\beta_1$/$\beta_2$), with some low intensity in between. The main difference observed 
between the  Ir$^{4+}$ (left panel) and Ir$^{5+}$ (right panel) compounds is the intensity ratio between these 
two features. For the Ir$^{4+}$ samples both features have a similar integrated intensity, while for the Ir$^{5+}$ 
samples the $\beta_1$/$\beta_2$ features are more intense.  

The fact that the presence of the different B cations (e.g.\ Mg, Y, Fe, Co, Ni) has little effect on the overall line shape
of the spectra, suggests that the spectra are dominated by the contribution from the iridium orbitals. Indeed, the tabulated
photo-ionization cross-section values for the Ir $5d$ orbitals are by far the largest in comparison to those of the transition 
metal $3d$ and the O $2p$ for x-rays with 6.5 keV energy as we have used in our HAXPES experiment
\cite{TRZHASKOVSKAYA200197,TRZHASKOVSKAYA2002257,TRZHASKOVSKAYA2006245}. We can in fact deduce now
already that the features $\alpha$ and $\beta_1$/$\beta_2$ are displaying the Ir $5d$ PDOS and that this Ir $5d$ PDOS
is rather similar across the set of compounds independent of the nature of the B cation, which by itself is quite remarkable.

\begin{figure*}
\centering
    \includegraphics[width=1.98\columnwidth]{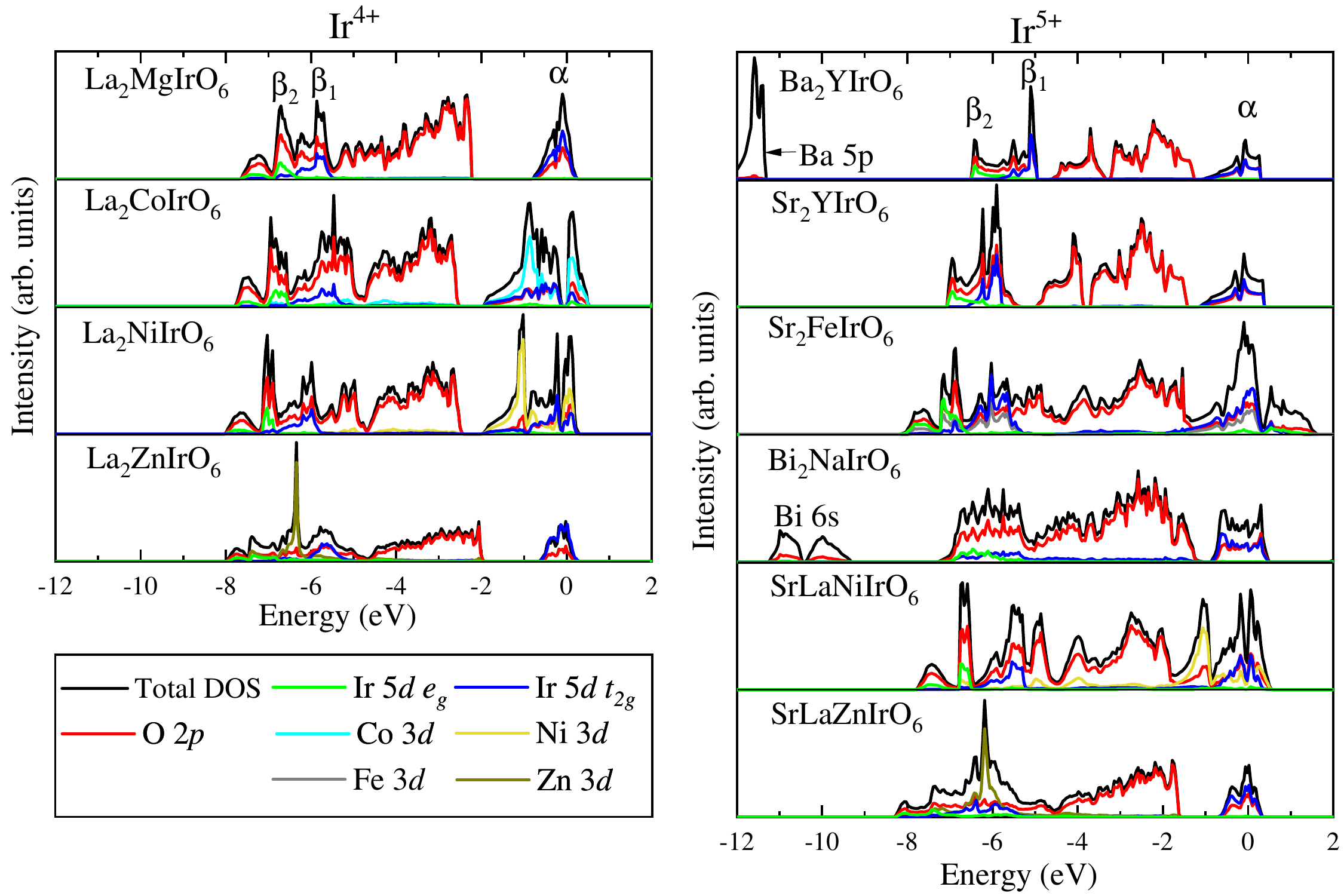}
    \caption{Calculated density of states (DOS) and the Ir $5d$ , O $2p$ and transition metal $3d$ partial density of 
     states (PDOS) of the double perovskite compounds with Ir$^{4+}$ (left panel) and Ir$^{5+}$ (right panel).}
    \label{fig_2}
\end{figure*}

In order to better understand the electronic structure of these iridates, we performed DFT calculations and
projected out the different orbital contributions to the valence band. Figure \ref{fig_2} shows the density of states 
(DOS) and the partial density of states (PDOS) of the Ir $5d$ $e_{g}$ and $t_{2g}$, O $2p$ as well as the $3d$ from the other transition 
metals on the B site. We can observe that the Ir $5d$ density is mostly located in the same regions as the main 
two features $\alpha$ and $\beta_1$/$\beta_2$ observed in the experimental spectra. The O $2p$ is present 
throughout the entire valence band, not only where the Ir $5d$ is present but also in the region between the two 
iridium features.

As for the contribution from the $3d$ orbitals, the intensity for the Co and Ni compounds is mostly slightly below the iridium states close to the Fermi energy, for the Fe compound it is more evenly distributed, while the Zn compound it is close to the deeper iridium states. { Here we note that the calculated energy position of the Zn $3d$ states deviates from the experiment. This is a short-coming of standard DFT calculations, and the inclusion of self-interaction effects is required to reproduce properly the spectra of ZnO and related materials \cite{Lim2012}.}
We also note that all DFT results produce a metallic state and that therefore electron correlations effects need to be included \cite{Pajskr2016} in order to reproduce a gap associated with the experimentally observed insulating behavior of the compounds.
{The purpose of our DFT calculations is to gain insight into single-particle processes like hybridization and crystal field splittings instead of reproducing the small bandgaps.}

\begin{figure*}
\centering
    \includegraphics[width=1.98\columnwidth]{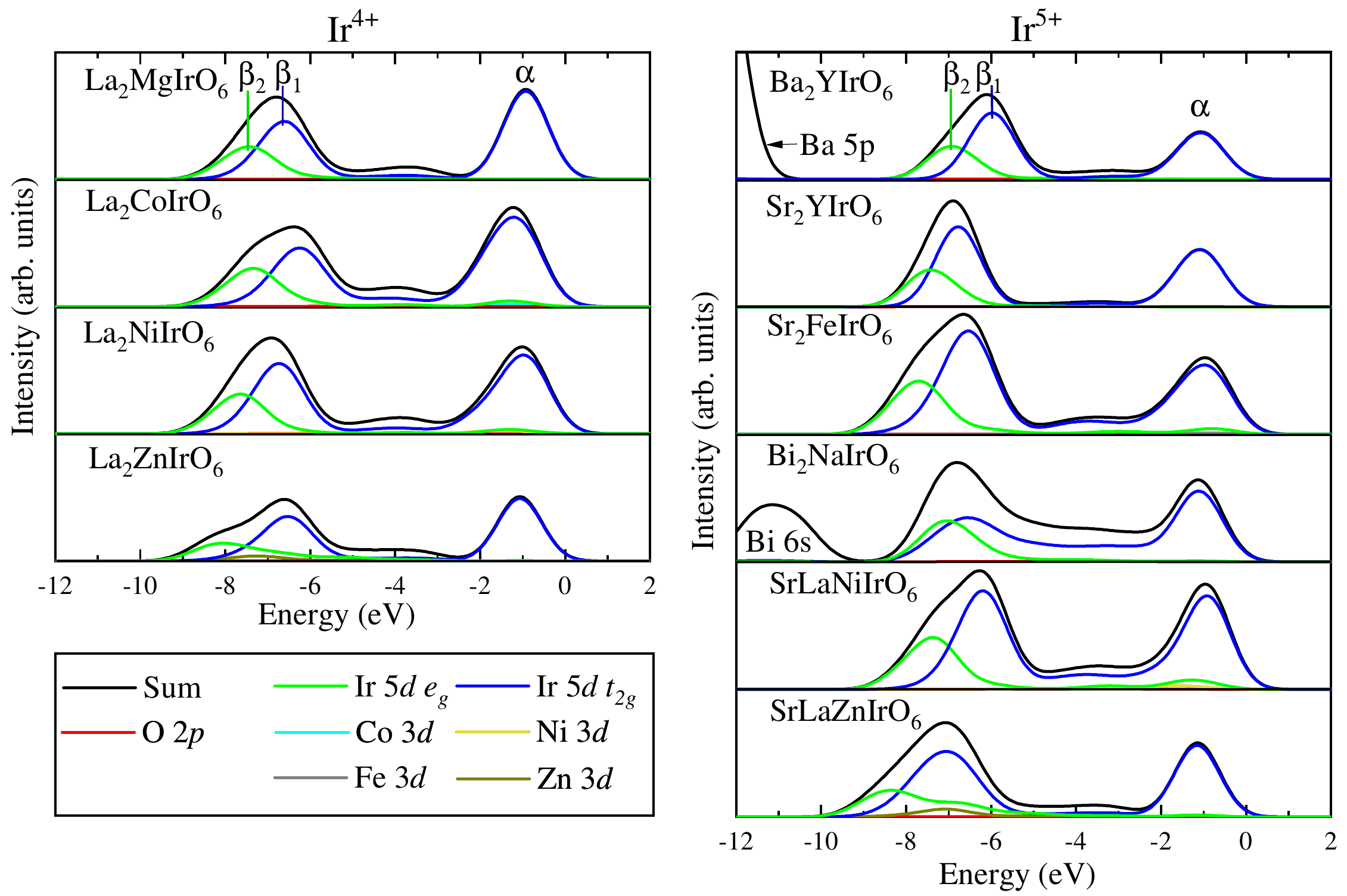}
    \caption{Calculated valence band spectra of the double perovskite compounds with Ir$^{4+}$ (left panel) 
    and Ir$^{5+}$ (right panel). The spectra were obtained by multiplying the calculated PDOS of the occupied 
    states by their respective photoionization cross sections at 6.5 keV photon energy, followed by a broadening 
    to account for experimental conditions, and then their summation. {Finally, a rigid shift is applied to simulate the gap.} }
    \label{fig_3}
\end{figure*}

In order to compare the DFT results to the experiment, we calculate the valence band spectra by multiplying each 
of the PDOS by their respective photoionization cross section at 6.5 keV photon energy as derived from Refs. \cite{TRZHASKOVSKAYA200197,TRZHASKOVSKAYA2002257,TRZHASKOVSKAYA2006245} and by the Fermi function 
to include only the contributions from the occupied states, followed by a broadening to account for the experimental 
resolution and their summation. This was done for all the PDOS included in the calculation (not only the ones shown 
in the figures). { Finally, the obtained spectra have been shifted rigidly to match the position of the experimental $\alpha$ feature and to simulate the experimentally observed gap.} The results are shown in Figure \ref{fig_3}.

We can notice first of all, that in all cases most of the intensity originates indeed from Ir 5d states. 
The contributions of O $2p$ and also of transition metal $3d$ orbitals are negligible in all cases. The remaining weight
can be attributed to states not explicitly represented in this plot, which are mostly $p$ states from 
Ba, La, Y, or Sr, which become much more relevant when measuring using hard x-rays \cite{Takegami2019}. 

Comparing the calculated with the experimental spectra we find good overall agreement.
There are some deviations when looking in more detail, for example, the shapes and positions of the features 
are not completely reproduced. Also, in some cases, the intensity in the region between the two main features 
is somewhat underestimated, which could be due to an underestimation of some of the cross-sections.
Nevertheless, it is safe to state that the overall experimental spectral features are well explained by the calculations.
In particular, we
observe that the materials trend for the intensity ratios of the two features $\alpha$ vs. 
$\beta_1$/$\beta_2$ when comparing the Ir$^{4+}$ and Ir$^{5+}$ set of compounds is well captured by the calculations.

\section{Analysis and Discussion}

In order to extract the dominant hopping parameters on a minimal basis, we perform Slater-Koster LCAO tight-binding
modelling on the experimental spectra. As representative compounds for this more detailed study we take La$_2$MgIrO$_6$ for Ir$^{4+}$ and Ba$_2$YIrO$_6$ for Ir$^{5+}$,
i.e., systems which do not contain $3d$ transition metal ions which otherwise could complicate the analysis due
to typically very strong correlation effects within the $3d$ shell.

To this end we start by numerical downfolding of the converged DFT Kohn-Sham bands to a Wannier orbital basis.
Besides Ir $5d$ and O $2p$ states we included also O $2s$ states. The reason for this inclusion is a
non-negligible hybridization between O $2s$ and the $5d$ $e_g$ states of iridium. If such hybridization effects
were included only implicitly, it would lead to a renormalized (enlarged) effective crystal field splitting by shifting
the Ir $5d$ $e_g$ states to higher energies (even before considering any hybridization effects with O $2p$ states).
Our choice for the explicit inclusion of O $2s$ states can be, hence, understood as the intention to stay as
close as possible to an atomic basis for our tight-binding analysis.

In our model, we distinguish two different types of O $2p$ orbitals, as hopping integrals vary 
depending on the symmetry of the corresponding bond. O $2p$ $\sigma$ orbitals, 
which are aligned along the Ir-O direction and hybridize mostly with the Ir $5d$ $e_{g}$ and O $2p$ $\pi$ orbitals,
which are aligned perpendicularly and hybridize mainly with the Ir $5d$ $t_{2g}$ states.

\begin{figure*}
\centering
    \includegraphics[width=1.98\columnwidth]{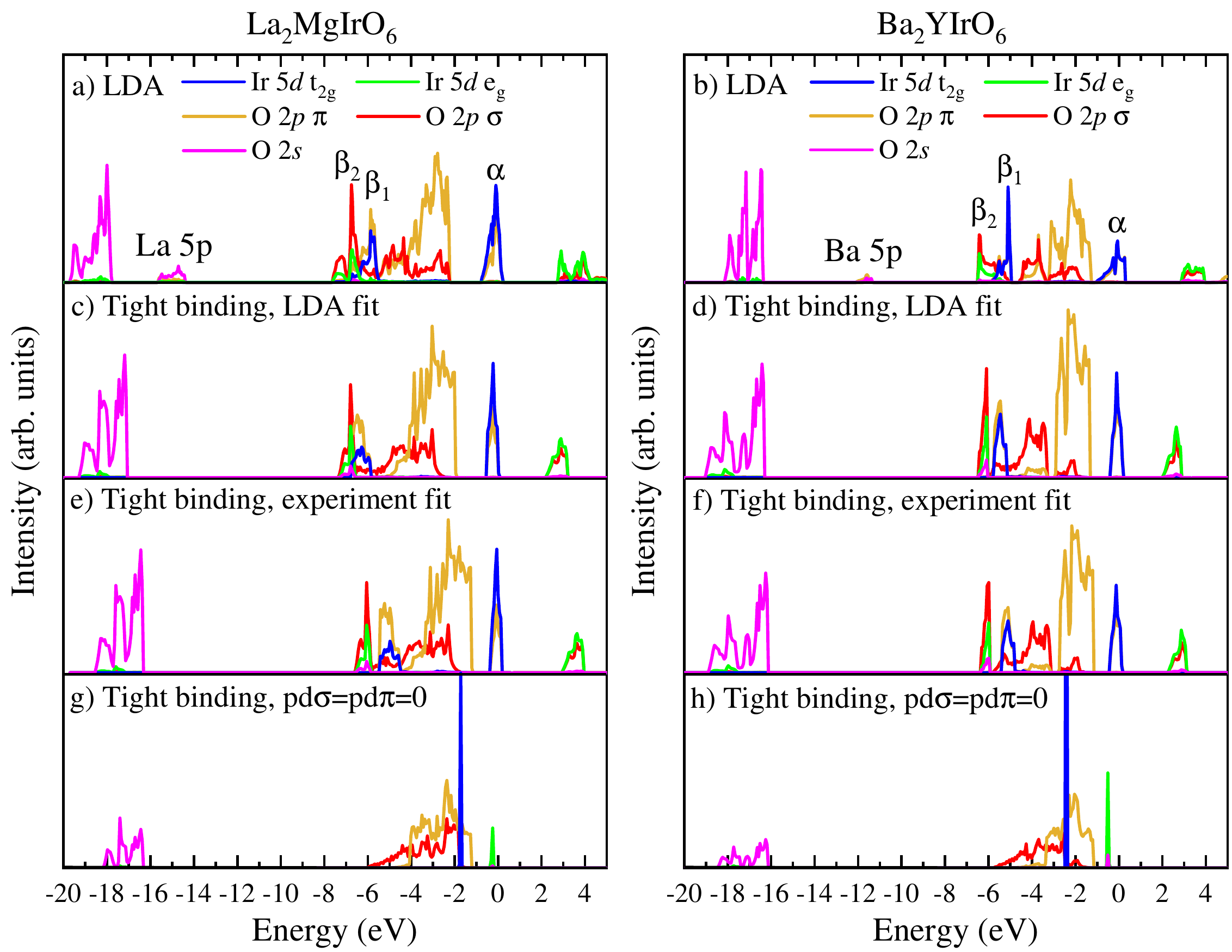}
    \caption{ 
     (a) and (b): Comparison of the DOS and PDOS obtained from the \textit{ab initio} calculations and the tight 
     binding calculations for La$_2$MgIrO$_6$ (left panel), chosen as a representative compound for the double 
     perovskites with Ir$^{4+}$, and Ba$_2$YIrO$_6$ (right panel), as representant for the double perovskites 
     with Ir$^{5+}$. O $2p$ $\pi$ ($\sigma$) corresponds to the contribution of the O $2p$ orbitals perpendicular 
     (parallel) to the Ir-O direction. 
     (c) and (d): Results of the tight binding calculations using the parameters obtained from the downfolding of the 
     \textit{ab initio} calculations. 
     (e) and (f): Results after fine tuning of the parameters to match the experimental spectra \cite{param_LMIO,param_BYIO}. 
     (g) and (h): Results of the tight binding calculations performed with the fine tuned parameters but with 
     $pd\pi=pd\sigma=0$, i.e., with the hybridization between Ir $5d$ and O $2p$ switched off. }
    \label{fig_4}
\end{figure*}

The DFT calculations with the projected Ir $5d$  $e_{g}$/$t_{2g}$ and O $2s$, $2p$ $\sigma$/$\pi$ PDOS are 
shown in Figs.\ \ref{fig_4} (a) and (b) for La$_2$MgIrO$_6$ and  Ba$_2$YIrO$_6$ respectively. For 
feature $\alpha$ at 1eV binding energy from the experiment (see Fig.\ \ref{fig_1}) we observe a clear Ir $5d$ $t_{2g}$ character. 
This is also true for feature $\beta_1$ at 6.2 eV.  Feature  $\beta_2$ at 7 eV, on the other hand, originates 
entirely from Ir $5d$ $e_g$ states. All features have appreciable O $2p$ $\pi$ and $\sigma$ character, respectively.

The results of the tight-binding calculations using the parameters as obtained from the downfolding
\cite{param_LMIO,param_BYIO} are displayed in panels (c) and (d). We observe that the tight-binding results 
reproduce the PDOS of Ir $5d$ found in DFT well. This implies that the most relevant hopping processes for the Ir $5d$ states are captured by our minimal tight-binding model. This also means the cations of the A and B sites do not play a significant direct role on the 
Ir $5d$ states. Starting from this set of parameters, we can adjust them in order to get a better match to the 
experimental spectra and thus obtain the parameters that describe best what we have observed in our measurements.

\begin{figure*}
\centering
    \includegraphics[width=1.98\columnwidth]{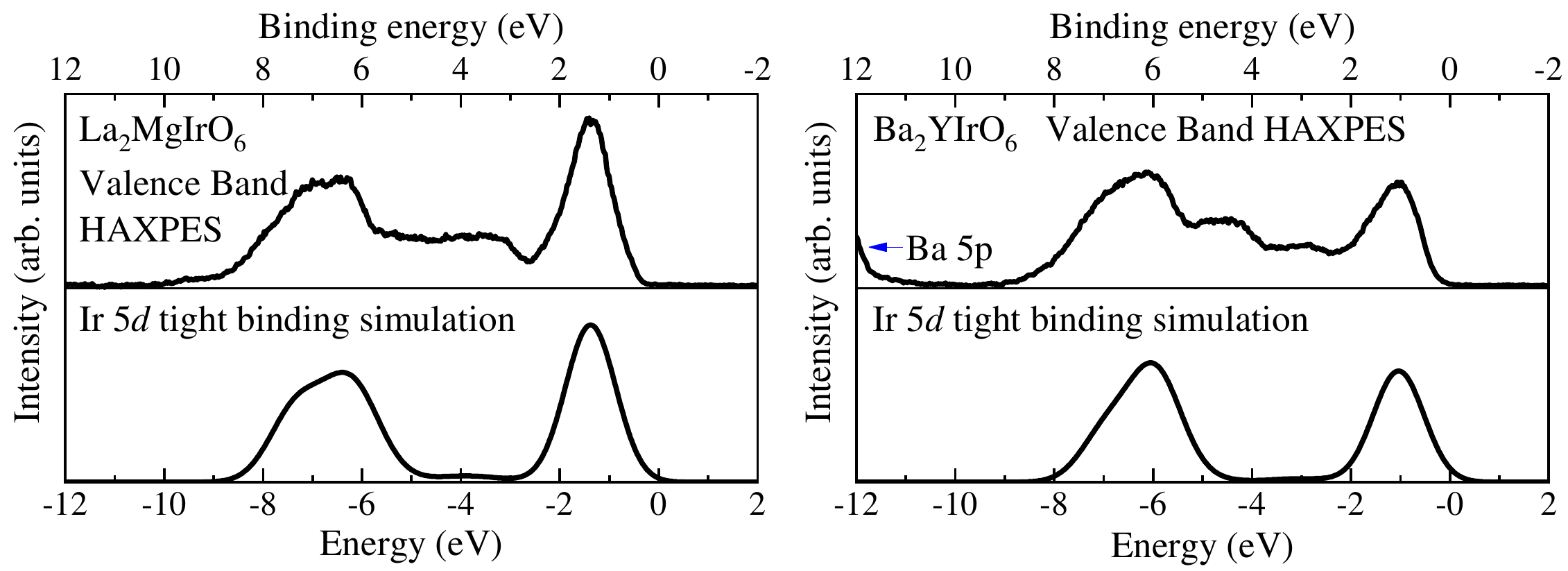}
    \caption{ Comparison of the experimental HAXPES spectra (top) with the simulated Ir $5d$ valence band 
    spectrum based on the optimized tight binding calculations (bottom) for La$_2$MgIrO$_6$ (left panel) 
    and Ba$_2$YIrO$_6$ (right panel).}
    \label{fig_5}
\end{figure*}

Figs.\ \ref{fig_4} (e) and (f) are the results after fine tuning the parameters \cite{param_LMIO,param_BYIO} in order to get the separation between features $\alpha$ and $\beta_1$/$\beta_2$ to match the experiment. Fig.\ \ref{fig_5} shows the comparison of the experimental spectra with the Ir $5d$ contribution obtained with 
this optimized set of tight-binding parameters. {The same procedure as described for the data in Fig.~\ref{fig_3} is used to simulate the experimental conditions.} The effect of the photoionization cross-sections is 
effectively already taken into account since we are looking into the Ir $5d$ only, and its contribution is by far the dominant 
one for the HAXPES spectra. We can observe that the Ir $5d$ PDOS replicates very well the features from the 
experiment. 

Finally, Figs.\ \ref{fig_4} (g) and (h) are the results of the calculations using the set of parameters for the experimental 
fitting but with pd$\pi =$ pd$\sigma = 0$, i.e., with the hybridization between the Ir $5d$ and O $2p$ switched off.
These plots reveal most clearly the effect of hybridization process between the Ir $5d$ and the O $2p$.
We can see that the Ir $5d$ $t_{2g}$ and $e_{g}$ states in (g) and (h) become delta-function like which implies that inter-site
iridium hopping is small and that most of the Ir $5d$ band formation is due to the hopping via the O $2p$ orbitals.
Most interesting is the energy position of the Ir $5d$ states when the hopping is set to zero: the Ir $5d$ $t_{2g}$ fall inside the O $2p$ $\pi$ band. 
For the Ir$^{4+}$ case (g), $t_{2g}$ is at the upper half of this oxygen band, and for the Ir$^{5+}$ (h), it is even 
in the middle. In other words, the O $2p$ $\pi$ and the Ir $5d$ $t_{2g}$ states are nearly degenerate in the Ir$^{4+}$ 
system and fully degenerate in the Ir$^{5+}$ material. 

With these findings we get a clearer picture of the hybridization process between the O $2p$ and the Ir $5d$ $t_{2g}$.  
As can be seen from Figs.\ \ref{fig_4} (e) and (f), bonding (feature $\beta_1$) - antibonding (feature $\alpha$) bands 
are formed with about 5 eV energy separation. This separation is much larger than the separation between the
O $2p$ $\pi$ band and the Ir $5d$ $t_{2g}$ states before the hybridization, which is about 2 eV for the  Ir$^{4+}$ systems 
and 0 eV for the  Ir$^{5+}$, as shown in Figs.\ \ref{fig_4} (g) and (h), respectively. The hopping integral is thus much
larger than the difference in the O $2p$ $\pi$ and Ir $5d$ $t_{2g}$ on-site energies. Consequently, the bonding and
antibonding states have highly mixed O $2p$ $\pi$ and Ir $t_{2g}$ characters. The simulations in Figs.\ \ref{fig_4} 
(e) and (f) show that the antibonding peak at 1 eV binding energy has a 60\% Ir $5d$ $t_{2g}$ character in the 
Ir$^{4+}$ systems and 50\% in the Ir$^{5+}$.

In the case of the $e_g$ states, they are positioned at higher energies (closer to the Fermi level)
than the  $t_{2g}$, and thus more separated from the O $2p$ $\sigma$ bands before hybridization, see 
Figs.\ \ref{fig_4} (g) and (h). Yet, even on the scale of this Ir $5d$ $e_g$ O $2p$ splitting, their $\sigma$ hopping integrals
are so large, that the energy separation between the resulting bonding (feature $\beta_2$) and anti-bonding (above the Fermi level)  
states reaches values of about 10 eV, see Figs.\ \ref{fig_4} (e) and (f). The hybridized states have a 50-50 
mixed character. In other words, also the Ir $5d$ $e_g$ and the O $2p$ $\sigma$ bands are effectively degenerate.

In developing models which include electron correlation effects and the spin-orbit entanglement explicitly in order to describe the (potentially) exotic magnetic properties, the degeneracy of the O $2p$ and Ir $5d$
states translates into a value for the O $2p$ to Ir $5d$ charge transfer energy which is essentially zero.
As a result, the assumption of an ionic $t_{2g}^{5}$ configuration in order to stabilize the pure $J_{eff}=1/2$ 
state for Kitaev physics can no longer be justified. There will be a substantial oxygen ligand hole character in the 
Ir $5d$ $t_{2g}$ Wannier orbitals and this extremely strong covalency must be taken into account in the quantitative 
evaluation of the models. This has far reaching consequences. Additional inter-site magnetic exchange interactions 
will be present, i.e.\ not only of the Kitaev type. Moreover, the exchange interactions will become very long-ranged 
\cite{Agrestini2017,Morrow2013}, leading also easily to anisotropies not foreseen in the Kitaev model. Nevertheless, it is still 
quite feasible to design materials that show a spin-liquid behavior, provided that the exchange interactions present 
can be tuned such that they compete and sufficiently cancel each other \cite{Winter2016, Winter2017, Kitagawa2018}.

\section{Conclusion}
We have measured the valence band of several A$_2$BIr$O_6$ and AA'BIrO$_6$ double perovskites containing 
either Ir$^{4+}$ or Ir$^{5+}$. The spectra display very strong similarities as far as the Ir $5d$ contribution
is concerned, pointing out common aspects in the O $2p$ and Ir $5d$ hybridization process. Density functional theory calculations and Slater-Koster LCAO tight-binding calculations provide a detailed explanation of the spectra and demonstrate
that the iridates are highly covalent systems with essentially zero O $2p$ to Ir $5d$ charge transfer energy. 
The consequence is that the exchange interactions become very long-ranged, thereby inhibiting the materialization of 
the pure Kitaev model. Nevertheless, it still would be possible to realize a spin-liquid system using the iridates 
with a proper tuning of the various competing exchange interactions.

\section*{Acknowledgements}
We would like to thank Y. H. Wu from the NSRRC for technical assistance during the HAXPES experiments.
Fruitful scientific discussions with M.~Vogl and D.~Mikhailova (both IFW Dresden) are gratefully acknowledged. The research in Dresden was partially supported by the Deutsche Forschungsgemeinschaft 
through SFB~1143 (project-id 247310070) and Grant No. 320571839. We acknowledge support 
from the Max Planck-POSTECH-Hsinchu Center for Complex Phase Materials. R. M. gratefully acknowledges support of the Humboldt Foundation.

\appendix
\section{Polarization dependence.}
\begin{figure}
\centering
    \includegraphics[width=0.999\columnwidth]{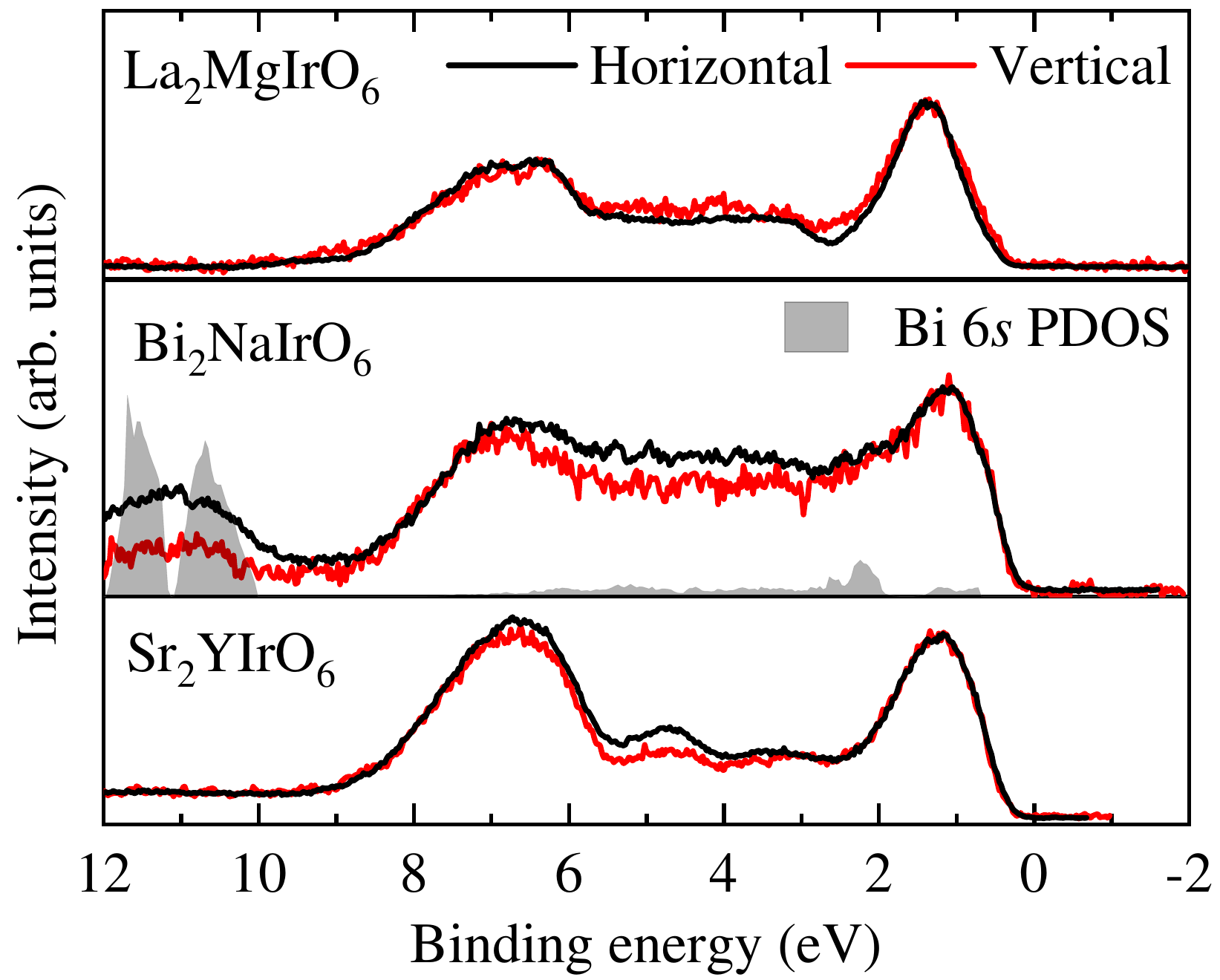}
    \caption{{Valence band HAXPES spectra ($h\nu=6.5$keV) of the double perovskite compounds La$_2$MgIrO$_6$, Bi$_2$NaIrO$_6$, and Sr$2$YIrO$_6$ taken with photoelectrons collected in the direction parallel (black, " Horizontal") and perpendicular (red," Vertical") to the electrical field vector of the photon beam. The calculated Bi $6s$ partial DOS is indicated for the Bi$_2$NaIrO$_6$ compound.}}
    \label{fig_7}
\end{figure}
{
Figure \ref{fig_7} displays the HAXPES valence band spectra of La$_2$MgIrO$_6$, Bi$_2$NaIrO$_6$, and Sr$_2$YIrO$_6$ taken with the photoelectrons collected in the direction parallel (black, “Horizontal”) and perperdicular (red, “Vertical”) to the electrical field vector of the photon beam.  The polarization dependence is relatively minor, except for the Bi containing iridate where a distinguishable suppression can be observed for the perpendicular direction in the $12$~eV-$10$~eV and $6$~eV–$3$~eV binding energy ranges. From HAXPES studies it is well known that the polarization dependence is the strongest for $s$-type orbitals since they have a $\beta$-asymmetry parameter close to 2 \cite{TRZHASKOVSKAYA200197,TRZHASKOVSKAYA2002257,TRZHASKOVSKAYA2006245,Weinen2015,Takegami2019}.  The observed suppression therefore matches very well the presence of the Bi $6s$ partial density of states as found from the band structure calculations.
}
\end{document}